\documentclass[aps,pra,twocolumn,floatfix,superscriptaddress]{revtex4-2}

\usepackage{amsmath}
\usepackage{amssymb}

\usepackage{subfigure}
\usepackage{graphicx}
\usepackage[T1]{fontenc}
\usepackage{xcolor}

\newcommand{\WUT}{\mbox{Faculty of Physics, Warsaw University of Technology, Ulica Koszykowa 75, PL-00662 Warsaw, Poland}}
\newcommand{\IFPAN}{\mbox{Institute of Physics, Polish Academy of Sciences, Aleja Lotnikow 32/46, PL-02668 Warsaw, Poland}}

\begin{document}

\title{Atomic boson-fermion mixtures in 1D box potentials: Few-body and mean-field many-body analyses}
\date{\today}

\author{Bishal Parajuli}
\affiliation{Department of Physics, University of California, Merced, CA 95343, USA.}

\author{Daniel P\k{e}cak}
\affiliation{\IFPAN}
\affiliation{\WUT}

\author{Chih-Chun Chien}
\email{cchien5@ucmerced.edu}
\affiliation{Department of Physics, University of California, Merced, CA 95343, USA.}

\begin{abstract}
We study binary atomic boson–fermion mixtures confined in one dimensional box potentials by few-body theory with contact interactions and mean-field many-body theory with density-density interactions. A variety of correlations and structures arise as the inter- and intra- species interactions are tuned. Both few-body and many-body results show that miscible phase and three-chunk phase separation are directly observable in the density profiles. Meanwhile, two-chunk phase separation can be inferred from the few-body correlations and many-body density profiles. We present phase diagrams of selected types of atomic mixtures to show where different structures survive. 
The few-body analysis demonstrates that two-body correlation functions can reveal information relevant to the results from many-body calculations or experiments. 
From the many-body density profiles in the phase-separation regime, we extract the healing lengths of each species and explain the scaling behavior by an energy-competition argument.
\end{abstract}

\maketitle

\section{Introduction}
Advancement in trapping and cooling atoms has made it possible to study quantum many-body physics using ultracold atoms. After the realizations of Bose-Einstein condensate (BEC) in a single-component atomic gas \cite{AndersonM.H1995OoBC,PhysRevLett.75.1687,PhysRevLett.75.3969}, experimental groups have created condensates in atomic boson-boson mixtures  \cite{PhysRevLett.78.586,phase-sep-exp} and BEC of molecules in two-component Fermi gases with tunable interactions  \cite{PhysRevLett.91.250401, GreinerMarkus2003Eoam}.
Those experiments explored two-component mixtures using two hyperfine states of the same species.
Later, fermionic atoms were mixed with bosonic atoms in several examples, including ${}^7$Li--${}^6$Li mixtures  \cite{doi:10.1126/science.1059318,PhysRevLett.87.080403}, ${}^{23}$Na--${}^6$Li mixtures  \cite{PhysRevLett.88.160401}, ${}^{87}$Rb--${}^{40}$K mixtures \cite{PhysRevLett.89.150403,PhysRevLett.96.020401,doi:10.1126/science.1077386}, ${}^{87}$Rb--${}^6$Li mixtures \cite{PhysRevA.77.010701}, ${}^{87}$Sr--${}^{84}$Sr mixtures \cite{PhysRevA.82.011608}, ${}^{41}$K--${}^{6}$Li mixtures \cite{PhysRevLett.117.145301}, and ${}^{133}$Cs--${}^6$Li mixtures \cite{PhysRevLett.119.233401}. In general, binary atomic boson-fermion mixtures with repulsive inter-species interactions demonstrate a bosonic BEC and a single-component normal Fermi gas since pairing mechanism is not involved. 

On the other hand, theoretical investigations have been carried out to characterize multi-component ultracold atomic systems. For example, the ground-state properties of two-component bosons confined between hard walls \cite{PhysRevA.79.033607}, phase separation in harmonically trapped and mass-imbalanced fermion-fermion mixtures\cite{JasonHo2013PhaseSeparation,Pecak}, one-dimensional harmonically trapped boson-fermion mixtures \cite{BellottiFilipeF.2017Cnaa}, ground-state densities of repulsive two-component Fermi gases \cite{PhysRevA.93.023612}, and repulsive boson-fermion mixtures in harmonic traps \cite{PhysRevLett.80.1804,PhysRevA.59.2974} have been studied, to name a few. It has been demonstrated that the structures of atomic mixtures depend on the inter- and intra- species interactions. For example, repulsion between bosons and fermions in a binary mixture leads to spatial separation, minimizing the overlapping region \cite{PhysRevLett.80.1804,PhysRevA.59.2974}. In contrast, attraction in a mixture may lead to collapse \cite{PhysRevLett.96.020401,doi:10.1126/science.1077386} or droplet formation \cite{rakshit2019self,karpiuk2020bistability}. There have been studies of thermodynamics and structural transitions of binary atomic boson-fermion mixtures using path integral formalism \cite{Tom18}. Ref. \cite{PhysRevA.100.063623} on binary boson-boson or fermion-fermion mixtures in 1D box potentials has suggested different phase-separation structures, including those induced by the mass imbalance. Many interesting phenomena of 1D atomic mixtures have been reviewed in Ref. \cite{2019SowinskiRPP}.
When compared to purely bosonic or fermionic gases, boson-fermion mixtures are particularly interesting as the two components follow different spin-statistics. For example, sympathetic cooling may use bosons to cool down fermions, as summarized in Ref. \cite{Onofrio_2016}.

Conventionally, ultracold atomic gases have been trapped in harmonic potentials, causing inhomogeneous density profiles. Recent progress in engineering optical potentials has brought us box potentials \cite{van_Es_2010,PhysRevLett.110.200406,ChomazLauriane2015Eocv,Tajik:19,PhysRevLett.118.123401,PhysRevLett.120.060402,PhysRevResearch.3.033013}, which simplify the comparison between theories \cite{das2003bose, fratini2012mass, manabe2019single} and experiments in the bulk. Homogeneous BEC of trapped bosonic atoms in quasi-1D \cite{van_Es_2010,Tajik:19}, 2D \cite{ChomazLauriane2015Eocv} and 3D \cite{PhysRevLett.110.200406}  have been realized. For two-component fermions, homogeneous 3D Fermi gases \cite{PhysRevLett.118.123401} and 2D Fermi gases \cite{PhysRevLett.120.060402} have been realized. Recently, dipolar dimers of non-reactive fermionic ${}^{23}$Na$^{40}$K molecules have been realized experimentally to analyze the collision of ultracold molecules in optical box potentials for a comparison with those in dipole traps \cite{PhysRevResearch.3.033013}. 

Here we envision future combinations of research on multi-component atomic gases and box potentials.
Explicitly, we study binary atomic boson-fermion mixtures in quasi-1D box potentials to explore the rich phase diagrams, ground state properties, and interface structures. The bosons can interact with themselves and with the fermions through two-body s-wave scattering, but identical fermionic atoms do not interact with each other due to Pauli exclusion principle that suppresses two-body s-wave scattering~\cite{pethick2008bose}. We will begin with the Hamiltonian and divide our analyses into two parts: Exact treatments of few-body systems and mean-field approximations of many-body systems. The methods are complementary and allow us to take a closer look at different regimes. The former reveals exact properties of the ground state and the correlations that explain the macroscopic picture. The latter gives access to the macroscopic structures, including those with broken-symmetry. The many-body picture is also typical in experiments probing single-particle properties, such as those measuring the density profiles. More sophisticated measurements could look within multi-particle correlations. Such state-of-the-art experiments are possible both in few-body \cite{Andrea2018Imaging, 2019BergschneiderNaturePhys, holten2022observation} and many-body \cite{haller2015single, GrossChristian2021Qgmf} regimes, and tools are known as atomic microscopes.

We will diagonalize the few-body Hamiltonian via the single-particle basis to obtain the ground state, though which the density profiles and two-body correlations can be evaluated. The complexity of the few-body calculation grows rapidly, calling for an approximate treatment for many-body systems.
By coupling the Gross-Pitaevskii equation of the bosonic condensate and Hartree approximation of the fermions with boson-fermion interactions, the many-body approximation will show a variety of density profiles in the ground state and maps out the phase diagrams for the most stable configuration. We then analyze effects of mass imbalance and interactions on the density profiles and two-body correlations. Both few-body and many-body results
show that the hard-wall boundary condition leads to structures different from those in a harmonic trap due to differences in the single-particle spectra. For example, the harmonic trap favors the core-plus-shell structure while a box potential can accommodate sandwich structures or two-chunk separation.

The rest of the paper is organized as follows. Sec.~\ref{sec:theory} summarizes the few-body and many-body formalisms of binary boson-fermion mixtures and the numerical procedures for simulations. Sec.~\ref{sec:fewbody} presents the phase diagrams, density profiles, and correlation functions from our few-body calculations. The analysis of the correlations will reveal more details of the structures than the density profiles. Sec.~\ref{sec:manybody} shows the phase diagrams and density profiles from the many-body mean-field calculations. By analyzing the widths of the density variations, we present the healing lengths of the bosons and fermions. A scaling argument from energy competitions captures the main features of the healing lengths. Sec.~\ref{sec:implications} discusses possible measurements of the phase-separation properties and implications for exotic phases of matter in atomic boson-fermion mixtures. Sec.~\ref{sec:conclusion} concludes our work.

\section{Theoretical framework}\label{sec:theory}
Here we summarize the theoretical frameworks of both few- and many -body pictures.
For the whole analysis, we assume equal population of the bosons and fermions with $N_b = N_f \equiv N$ confined in a quasi-1D box of length $L$. Depending on the picture considered, $N$ will be of the order of one or a hundred.
We denote the masses of the bosons and fermions as $m_b$ and $m_f$, respectively.

For a binary boson-fermion mixture, there are two coupling constants from the two-body s-wave collisions: The intraspecies interactions between bosons, $g_{bb}$, and the interspecies interaction between bosons and fermions, $g_{bf}$. For single-component fermions, the Pauli exclusion principle suppresses two-body s-wave collisions between identical fermions, hence $g_{ff}=0$.
The non-vanishing coupling constants $g_{bb}$ and $g_{bf}$ can be respectively expressed in terms of the two-body s-wave scattering lengths $a_{bb}$ and $a_{bf}$ away from resonance by \cite{pethick2008bose}
\begin{equation}
g^{3D}_{b\alpha}=2\pi\hbar^2a^{3D}_{b\alpha}/m_{b\alpha}
\end{equation}
with the reduced mass $m_{b\alpha}=(1/m_{b}+1/m_{\alpha})^{-1}$, where $\alpha \in \{b,f\}$.
Here $a^{3D}_{b\alpha}>0$ (or $<0$) corresponds to a repulsive (or attractive) interaction. In experiments, a magnetic field induced Feshbach resonance can be utilized to tune the 3D s-wave scattering length $a^{3D}_{b\alpha}$, leading to tunable interactions \cite{RevModPhys.82.1225,pethick2008bose}. In experiments, $g_{bb}$ and $g_{bf}$ may change together with the external magnetic field. Here we assume the two coupling constants can be tuned independently and map out the phase diagrams for selected atomic mixtures.

Quasi one-dimensional atomic gases can be realized by freezing the motion (with a tight confinement) in the transverse directions. Away from resonance, the coupling constant $g_{1D}$ of a 1D atomic gas can be expressed in terms of $a_{3D}$ via \cite{PhysRevLett.81.938}
$g^{1D}_{b\alpha}=\left( 2\hbar^2 a^{3D}_{b\alpha} \right)/ a_{\perp}^2$,
where $a_\perp$ is the length scale associated with the tight confinement in the transverse directions. Hence, $g^{1D}_{b\alpha}$ can be tuned by adjusting the ratio between $a^{3D}_{b\alpha}$ and $a_\perp$. In the following, we will drop the superscript $1D$ in the coupling constants and introduce the dimensionless parameters $\tilde{g}_{b\alpha}$ to rewrite the coupling constants as $g_{b\alpha}=\tilde{g}_{b\alpha} E_f^0/k_f^0$. Here $E_f^0$ is the Fermi energy and $k_f^0=\pi N_f/2L$ is Fermi wave-vector of a 1D noninteracting Fermi gas with the same particle number as the fermions of the mixture in the same 1D box potential. 

\subsection{Few-body theory}
Here we consider a mixture of equal numbers of bosons and fermions confined in a box of length $L$. 
The few-body Hamiltonian of the Bose-Fermi mixture reads:
\begin{equation}\label{eq:ham}
\begin{split}
 {\cal H} =& \sum_\alpha  \int_0^L dx \hat\Psi^\dag_\alpha(x) \left( -\frac{\hbar^2}{2m_\alpha} \frac{d^2}{dx^2} \right) \hat\Psi_\alpha(x)  \\
    +& \frac{g_{bb}}{2} \int_0^L dx \hat\Psi^\dag_b(x) \hat\Psi^\dag_b(x) \hat\Psi_b(x) \hat\Psi_b(x) \\
    +&       g_{bf}     \int_0^L dx \hat\Psi^\dag_b(x) \hat\Psi^\dag_f(x) \hat\Psi_f(x) \hat\Psi_b(x),
\end{split}
\end{equation}
where $\alpha \in \{b,f\}$ denotes bosons and fermions, respectively. The consecutive lines represent the single-particle Hamiltonian consisting only of the kinetic energy, the intrabosonic interactions, and the inter-species interactions between the bosons and fermions. 
The bosonic (fermionic) field operator $\hat\Psi_b(x)$ ($\hat\Psi_f(x)$) annihilates a particle at position $x$. The operators obey appropriate (anti)commutation relations:
\begin{subequations}\label{eq:commutation}
 \begin{align}
   [\hat\Psi_b^{\dagger}(x), \hat\Psi_b(x') ] &= \delta(x-x'),\\
  \{\hat\Psi_f^{\dagger}(x), \hat\Psi_f(x')\} &= \delta(x-x'), \\
   [\hat\Psi_b(x), \hat\Psi_f(x') ] &= 0. \label{eq:BFcommutation}
 \end{align}
\end{subequations}
Note that the last commutation relation, Eq.~\eqref{eq:BFcommutation}, is defined for distinguishable particles. It means that our choice of bosonic commutation relation is arbitrary as long as it is employed systematically throughout the analysis~\cite{Weinberg}, and the choice of the fermionic anticommutation relation would not alter the results.

The field operators can be expanded in a single-particle basis $\phi_n(x)$ as
\begin{equation}\label{eq:fieldoperator}
 \hat\Psi_\alpha(x) = \sum_n \phi_n(x) \hat a_{\alpha n},
\end{equation}
where $n$ runs over the complete basis spanned by $\{\phi_n(x)\}$, and the operator $\hat a_{\alpha n}$ annihilates a particle of type $\alpha$ in state $n$.
The single-particle basis $\phi_n(x)$ is the same for both species in the box:
\begin{equation}
 \phi_n(x) = \sqrt{\frac{2}{L}} \sin\left( \frac{n\pi}{L} x \right)
\end{equation}
and does not depend on the mass of the particles. In contrast, the single-particle energy depends on the inverse of the mass:
\begin{equation}\label{eq:spEnergy}
 E_{\alpha n} = \frac{\hbar^2 \pi^2 n^2}{2 L^2 m_\alpha}.
\end{equation}
Moreover, the number of particles is conserved in the few-body calculations, therefore the Hamiltonian Eq.~\eqref{eq:ham} can be diagonalized independently in each subspace of fixed $N_b, N_f$.

\subsubsection{Observables}
Similar to our previous study of atomic Bose-Bose and Fermi-Fermi mixtures~\cite{PhysRevA.100.063623}, we are mainly interested in the correlations and structures of the boson-fermion mixtures. 
The single-particle density of species $\alpha$ for the ground state $|\Phi_0 \rangle$ reads
\begin{equation}
 \rho_\alpha(x) = \langle \Phi_0 | 
 \hat\Psi^\dag_\alpha(x)
\hat\Psi^{\phantom\dag}_\alpha(x)
 | \Phi_0  \rangle.
\end{equation}
To quantify the homogeneity of the system, we introduce the following definition of homogeneity:
\begin{equation}\label{eq:Homo}
\mathfrak{h}=1-\int_0^L \frac{|\rho_b(x)-\rho_f(x)|}{N_b+N_f} dx.
\end{equation}
Here the definition by construction gives $\mathfrak{h}=0$ for the miscible (homogeneous) phase and $\mathfrak{h}=1$ for total phase separation with no overlap between the densities of the two species. We mention that other indicators, such as a weighted sum of the entropy of mixing or entropy of localization~\cite{Richaud19}, have been introduced to characterize the structures of boson-boson mixtures.

The homogeneity gives an important information about the structure but does not tell the whole story. First of all, the above consideration concerns the ground-state properties. However, a ground state can be in a superposition, which cannot be observed directly via single-particle measurements in experiments. After a measurement, the wave function collapses, and one realization is observed. 
Therefore, to shed light on the underlying structures of boson-fermion mixtures, we analyze two-body correlations. In particular, we will focus on the density-density correlations in real space from the ground state $|\Phi_0 \rangle$, defined by
\begin{subequations}
\begin{align}
 C_{bb}(x,y) &= \langle \Phi_0 | \hat\Psi^\dag_b(x) \hat\Psi^\dag_b(y)
\hat\Psi^{\phantom\dag}_b(y) \hat\Psi^{\phantom\dag}_b(x) | \Phi_0  \rangle, \\
 C_{ff}(x,y) &= \langle \Phi_0 | \hat\Psi^\dag_f(x) \hat\Psi^\dag_f(y)
\hat\Psi^{\phantom\dag}_f(y) \hat\Psi^{\phantom\dag}_f(x) | \Phi_0  \rangle, \\
 C_{bf}(x,y) &= \langle \Phi_0 | \hat\Psi^\dag_b(x) \hat\Psi^\dag_f(y)
\hat\Psi^{\phantom\dag}_f(y) \hat\Psi^{\phantom\dag}_b(x) | \Phi_0  \rangle.
\end{align} 
\end{subequations}
Here $x,y$ denote two points inside the 1D box.

\subsubsection{Numerical calculation}
\begin{figure}[ht]
\includegraphics[width=0.49\textwidth]{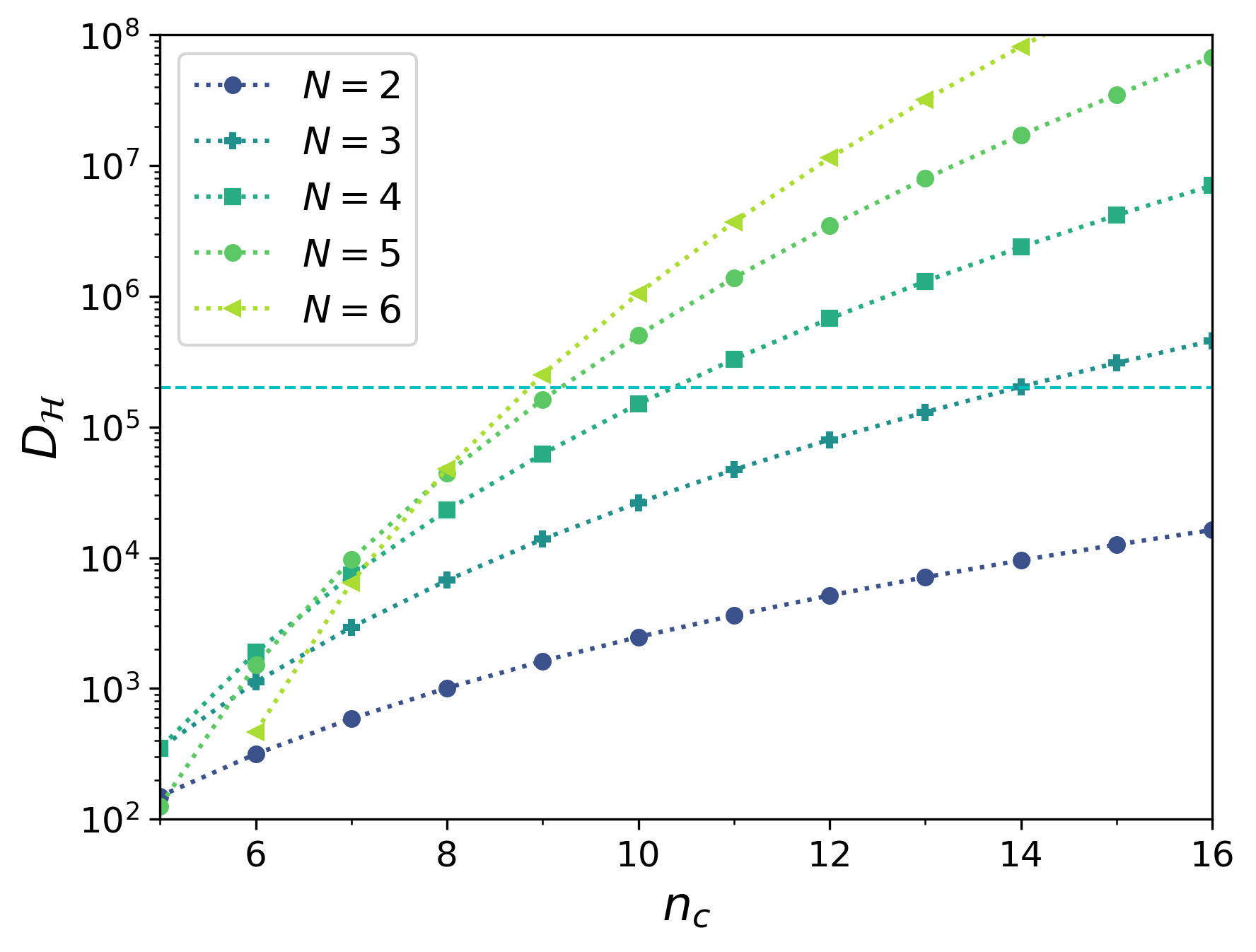}
\caption{\label{fig:fockdim}
The dimension $D_{\cal H}$ of the Hilbert space ${\cal H}$ of a mixture with $N$ bosons and $N$ fermions with a cutoff parameter $n_c$. The exponential growth is very rapid in the limit of large $n_c$. The dashed horizontal line shows the maximal dimension considered in the following studies.}
\end{figure}

The formalism introduced in Eq.~\eqref{eq:fieldoperator} assumes an infinite sum. In practice, to calculate the properties of desired few-body states, one introduces a cutoff in the number of single-particle orbitals used. With such a numerical approximation, the matrix elements of the Hamiltonian \eqref{eq:ham} are calculated. It is worth noting that the dimension $D_{\cal H}$ of the Hilbert space ${\cal H}$ grows exponentially with $n$, as shown in Fig.~\ref{fig:fockdim}. 

In our calculations, the dimension $D_{\cal H}$ is around $2\times 10^{5}$, and the results do not change significantly with increasing $n$. We note that the Hamiltonian is an operator and has dimension $D_{\cal H}^2$. However, the Hamiltonian matrix for the boson-fermion mixture studied here is sparse. Therefore, we can very effectively use the Arnoldi package~\cite{ARPACK1998Sorensen} to perform exact diagonalization and obtain the lowest-energy eigenstate. Since we use the method of exact diagonalization extensively for many years, the details can be found, for example in Ref. \cite{PhysRevA.100.063623} or in the Appendix of Ref. \cite{pecak2022unconventional}.
For each given $N=4$, mass ratio $m_b/m_f$, and cutoff $n_c=10$, we performed $100$  diagonalizations to obtain the $\tilde{g}_{bb}$-$\tilde{g}_{bf}$ phase diagram.

It is worth noting that there is a crucial difference between the boson-fermion mixtures and our previous studies of Bose-Bose mixtures and Fermi-Fermi mixtures~\cite{PhysRevA.100.063623}. 
When both components obey the same spin-statistics, the only difference in the energy scale in the single-particle picture comes from the relation Eq.~\eqref{eq:spEnergy}, i.e., the mass dependence. For boson-fermion mixtures, however, there is another difference at the single-particle level: The Fermi energy of the fermions differs significantly from the last level occupied by the bosons. The difference in the energy scales leads to more demanding overhead in the calculations.

\subsection{Mean-field many-body theory}
In a mean-field treatment of many-body systems, the contact interaction is coarse-grained into a density-density interaction~\cite{pethick2008bose}, which ignores the details of the wavefunction and only accounts for the energy change due to the overlap of the density profiles. 
The ground-state energy functional $E[\psi_b,\psi_{f,1},\cdots,\psi_{f,N_f}]$ of a binary boson-fermion mixture in a 1D box potential can be written as
\begin{eqnarray}\label{eq:EFun}
   E&=& \int_0^Ldx \Big[\frac{\hbar^2}{2m_b}N_b|\partial_x\psi_b|^2+\frac{\hbar^2}{2m_f}\sum_{i\leq N_f}|\partial_x\psi_{f,i}|^2\nonumber \\
    &&+\frac{1}{2}g_{bb}N_b^2|\psi_b|^4+g_{bf}N_b|\psi_b|^2\sum_{i\leq N_f}|\psi_{f,i}|^2\Big].
\end{eqnarray}
Here $\sqrt{N_b}\psi_b$ is the condensate wavefunction and $\psi_{f,i}$ is the $i$th fermionic eigen-state. The normalization conditions $\int_0^L dx|\psi_b|^2=1$ and $\int_0^L dx|\psi_{f,i}|^2=1$ for all $i$ are imposed.

In the mean-field description of the ground state, the condensate wavefunction describing the bosons is governed by the Gross-Pitaevskii equation~\cite{pethick2008bose,pitaevskii2003bose}. To find the minimal-energy configuration, we implement the imaginary-time formalism~\cite{Fetter_book,pethick2008bose} by searching for the stable solution to the imaginary-time evolution equation $-\partial \psi_b/\partial \tau=\delta E/\delta \psi_b^*$ in the $\tau\rightarrow \infty$ limit, starting from a trial initial configuration.  The normalization $\int|\psi_b|^2 dx =1$ is imposed at each imaginary-time increment to project out higher-energy states. Here $\tau=it$ is the imaginary time. Explicitly,
\begin{equation}\label{gpeqn1}
       -\hbar\frac{\partial \psi_b}{\partial \tau} = -\frac{\hbar^2}{2m_b}\partial_x^2\psi_b +   g_{bb}\rho_b\psi_b
      + g_{bf}\rho_f\psi_b,
\end{equation}
where, $\rho_{\alpha}$ with $\alpha \in \{b,f\}$ denotes the bosonic and fermionic density, respectively. 
Meanwhile, we describe the fermions by using the Hartree approximation, which leads to set of eigenvalue equations:
\begin{equation}
    -\frac{\hbar^2}{2m_f}\frac{\partial^2\psi_{f,i}}{\partial x^2} + g_{bf}\rho_b\psi_{f,i}= E_i\psi_{f,i}.
\end{equation}
We choose the units so that $\hbar=2m_f=1$. 
The coupled equations of the bosons and fermions are then solved together to obtain a configuration for the boson-fermion mixture.

The density profiles can be obtained from the condensate wavefunction and fermion wavefunctions via
\begin{equation}
    \rho_b=N_b|\psi_b|^2,~
    \rho_f=\sum_{i\leq N_f}|\psi_{f,i}|^2.
\end{equation}
The total number of particles of each species is given by
\begin{equation}
N_{\alpha}=\int_0^L dx \rho_\alpha.
\end{equation}
It is possible to obtain different solutions from different initial conditions that respect or violate the parity symmetry. In our numerical calculations, we have tried as many different initial states as possible and collected their final solutions. By comparing the ground-state energies via Eq.~\eqref{eq:EFun} from those different solutions, the lowest-energy state can be identified.

\begin{figure*}[th]
\includegraphics[width=0.99\textwidth]{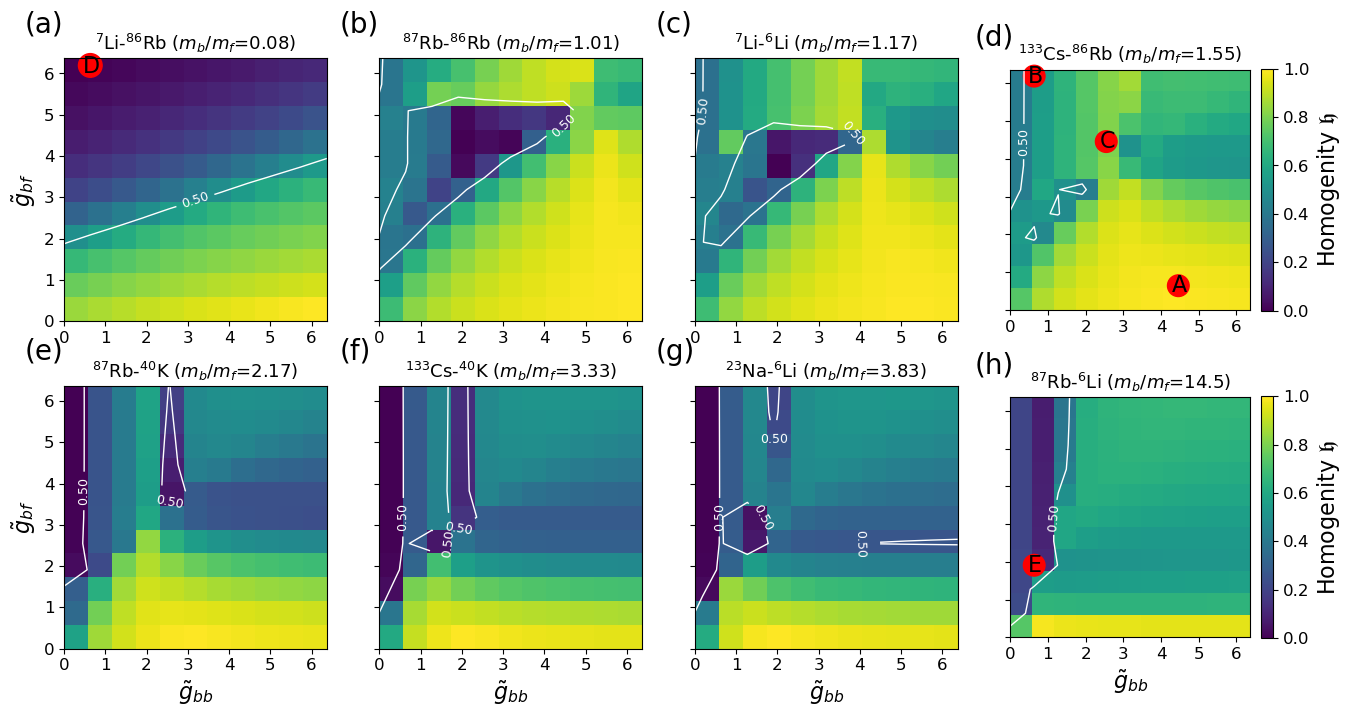}
\caption{\label{fig:miscibility}
The $\tilde{g}_{bb}$-$\tilde{g}_{bf}$ phase diagrams for various boson-fermion mixtures with $N_b=N_f=N=4$, showing the homogeneity $\mathfrak{h}$ of different mixtures with increasing mass ratio $m_b/m_f$. We smoothed the results obtained for interactions changing by $\Delta \tilde{g}_{b\alpha}=0.64$. While the diagrams becomes featureless for large mass imbalances, in the region of $m_b/m_f\approx 1.55$, the structures are very rich. The yellow regions ($\mathfrak{h}\approx 1$) mark a miscible mixture while the blue region ($\mathfrak{h}\approx 0$) shows phase separation. Here the quantities $\mathfrak{h}$, $\tilde{g}_{bb}$, and $\tilde{g}_{bf}$ are dimensionless.
}
\end{figure*}

\begin{figure*}[th]
\includegraphics[width=0.99\textwidth]{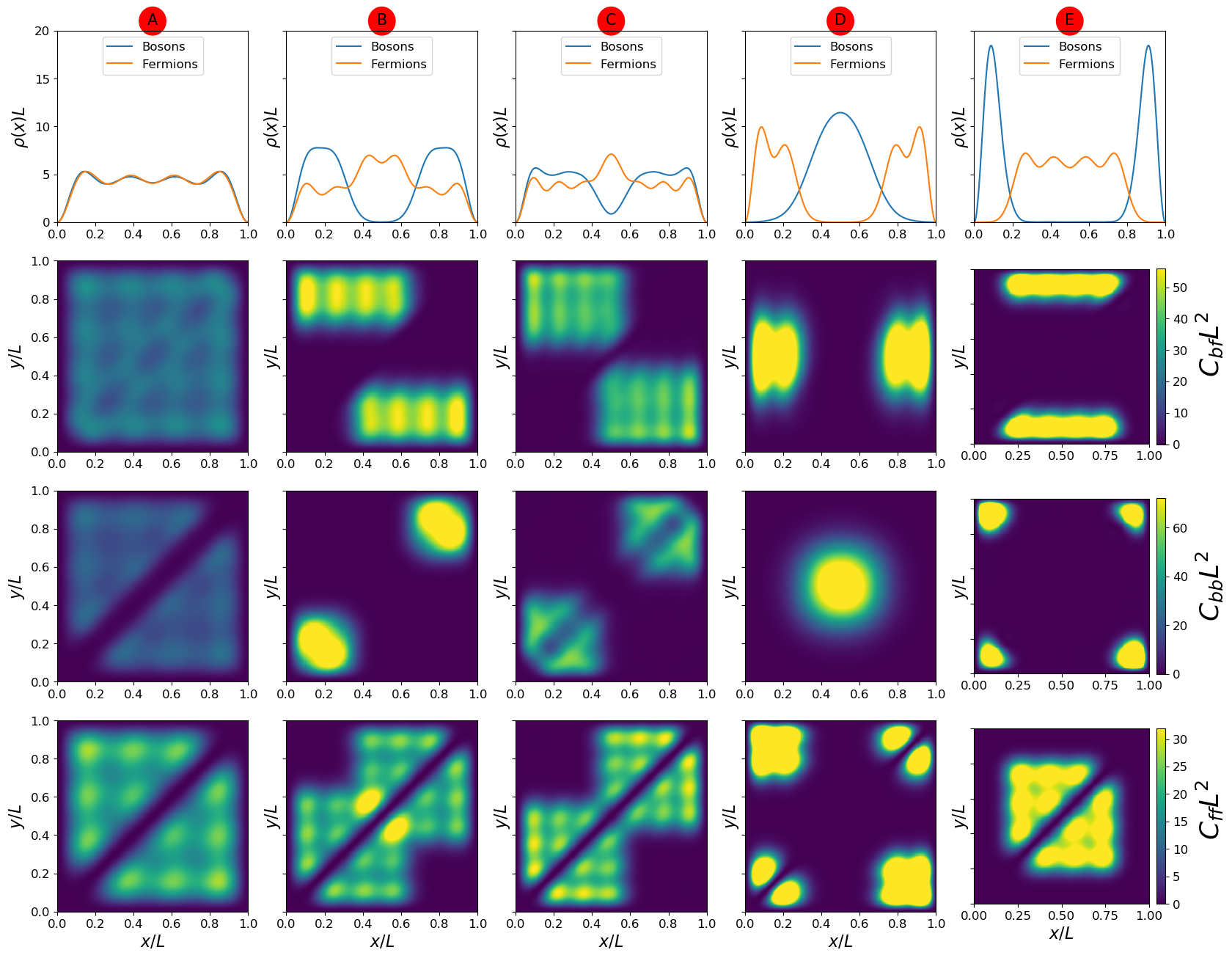}
\caption{\label{fig:states}
Top row: Density profiles, $\rho_\alpha(x)$. Second to the last rows: Two-body correlations $C_{bf}(x,y)$, $C_{bb}(x,y)$, and $C_{ff}(x,y)$ for the corresponding mixtures, respectively. Here $N_b=N=N_f=4$. From the left column to the right, we present the results from the points labeled on Fig.~\ref{fig:miscibility}. Here the two-body correlations inform case A as miscible, cases B and C as 2-chunk separation, and cases D and E as 3-chunk separation. }
\end{figure*}

\section{Few-body results}\label{sec:fewbody}
The boson-fermion mixtures have multiple parameters, implying rich structures and phenomena. Moreover, these parameters are not merely theoretical variables but associated with quantities that can be controlled in experiments either by choosing atomic species to vary the masses $m_\alpha$ or by tuning the inter- and intra- species interactions $\tilde{g}_{b\alpha}$. 

\subsection{Limiting cases}
Before focusing on particular results, we consider some limits of the parameters and simple scenarios to give us physical intuition.
First of all, when $\tilde{g}_{bf}=0$, the two species are independent of each other, so they can be treated on their own. 
Especially, for strongly interacting bosons (i.e., for $\tilde{g}_{bb} \rightarrow \infty$,) the density profile of the bosons tends to that of non-interacting fermions. Due to the equal numbers of both species, the density profiles will overlap perfectly, giving rise to a homogeneous mixture. Note that in this limit, the mass ratio is irrelevant since the bosons and fermions are decoupled. Therefore, the bottom of the $\tilde{g}_{bb}-\tilde{g}_{bf}$ phase diagrams are similar for different types of mixtures. As the inter-species interaction $\tilde{g}_{bf}$ increases, the differences start to come in while the homogeneity is lost. Those features are clearly shown in Fig.~\ref{fig:miscibility}.

In contrast, by setting $\tilde{g}_{bb}=0$, the only interaction is between the bosons and fermions, which should favor phase separation. This is because the bosons are noninteracting and can be described by just the lowest single-particle orbital, where the bosons bunch together because of their spin-statistics.
Due to the parity symmetry of the box, one species will stay in the middle, and the other is divided into two parts on the two sides. Since the Pauli exclusion principle favors separation of fermions, the rule of thumb  suggests that the bosons will stay in the middle while the fermions extend to the two sides. That is the case under the assumption that both species have comparable kinetic energies.
However, here the kinetic energy strongly depends on the mass. In particular, by increasing the mass of one species, the single-particle energy goes down as $1/m_\alpha$, favoring the species for accommodating spatial distortion and staying near the walls. In lithium-rubidium mixtures, the masses differ by an order of magnitude. It means in practice, for sufficiently large $\tilde{g}_{bf}$, the lithium atoms will stay in the middle and the rubidium atoms will spread towards the walls. The miscible-immiscible transition occurs sharply in a narrow regime, see Fig.~\ref{fig:miscibility}(a) and Fig.~\ref{fig:miscibility}(h).
For a huge mass imbalance, the kinetic energy plays an important role in determining the structure.

We remark that the main features of those results are due to the confinement of the box. Moreover, the fermion density is strongly modulated due to the Pauli exclusion principle, which alters how the components mix and separate. In the many-body treatment, those features are still observable. We also note that particular features may emerge in the few-body results. For instance, one can observe 'islands' of homogeneity in the regime of phase separation in Fig.~\ref{fig:miscibility}(b)-(d).
Those particular features stem from the subtle interplay between the interactions, spin-statistics, and the kinetic energies that strongly depend on the masses.

\subsection{Miscible and two-chunk structure}
As a concrete example, we first focus on the $^{133}$Cs-$^{86}$Rb mixture with a mass ratio $m_b/m_f\approx 1.55$. We choose five representative states with $(\tilde{g}_{bb}, \tilde{g}_{bf})$ given by A: (4.46, 0.64), B: (0.64, 6.37), C: (2.55, 4.46), 
for various interaction strengths (see Fig.~\ref{fig:miscibility}(d)). 
In Fig.~\ref{fig:states}, we analyze single-particle and two-body properties for those states. In the consecutive rows we present the densities $\rho_\alpha$ and the correlations $C_{bf}, C_{bb},$ and $C_{f\!f}$. 

Since the bosons are heavier, they stay near the walls of the box. Moreover, when $x=y$ in the plots of the correlations, the corresponding values are zero because of either the repulsion between bosons or the Pauli exclusion principle of the fermions. 
As we remarked earlier, when the intraspecies interaction dominates like the bottom of Fig.~\ref{fig:miscibility}(d), the system is a miscible mixture, as one can see in Fig.~\ref{fig:states}A. Apart from the fact that two identical fermions would avoid each other  due to the vanishing correlation at the same position, the probability of measuring two particles is almost uniform. This property fully fulfills the definition of a miscible mixture.
Furthermore, no clear separation is seen in the weakly interspecies interaction regime. The slight distortion near the hard walls is due to the boundary condition. With the size of the box goes to infinity, the homogeneity is expected to approach one for the miscible phase. 

Next, we observe that the density profiles in the stronger interspecies interaction regimes, as shown in the first row of Fig.~\ref{fig:states}, separate into three parts  with the lighter species in the middle. For a finite system in a box, the separation is not sharp unless very strong  interactions are involved. 
However, we have to be careful when interpreting the single-particle density profiles because there are superpositions of states respecting the parity symmetry, which in 1D is equivalent to the mirror symmetry with respect to the center of the box. As will be shown in the mean-field approach and expected in experiments, the parity symmetry may be broken by, for example, imperfections in the preparation and/or trap potential, fluctuations from the environment, or rounding in numerical evaluation. If the parity symmetry is broken, two-chunk structures in the density profiles may emerge with the bosons and fermions occupying opposite regions of the box to lower the interaction energy. 
To further differentiate the phase-separation structures in the few-body results, we analyze the two-particle correlations between the atoms.

For cases B and C of Fig. \ref{fig:states}, 
the tendency to form two-chuck structures if the parity symmetry is broken can actually be observed in the two-body correlations. First of all, the inter-species correlation 
$C_{bf}$ shows that the existence of a boson on one side of the box (left or right) corresponds to the existence of a fermion on the opposite side. Meanwhile, all the bosons are seen to congregate on one side according to $C_{bb}$. Similarly, all the fermions can be found either on the left or right half of the box according to $C_{f\!f}$. The behavior of fermions is a bit counter-intuitive since it implies the Pauli principle is not enforced. A careful analysis shows that two fermions are correlated in the left or right part of the box while the Pauli exclusion principle prohibits identical fermions from occupying the same position, which is reflected by the vanishing correlation along the diagonal of the plot. 

Similar effects are also observable in the two-particle correlations shown in panel C of Fig. \ref{fig:states} for stronger repulsion between the bosons.
The two-body correlation landscape showing aggregations of same species and separation of different species explains that the underlying structure is actually two-chunk, but the superposition in the ground state conceals it in the single-particle density profile. We anticipate that two-chunk phase separation will be revealed in the many-body limit after the parity symmetry is broken.

\subsection{Three-chunk structure}
Cases D and E of Fig. \ref{fig:states} show features that are associated with three-chunk structures of $^{7}$Li-$^{86}$Rb mixtures with $m_b/m_f=0.08$ and $^{87}$Rb-$^{6}$Li mixtures with $m_b/m_f=14.5$, respectively, see also Fig.~\ref{fig:miscibility}(a) and (h). Again, the few-body single-particle density profiles may or may not reflect the many-body structures after the parity symmetry is broken. It is thus crucial to analyze the correlations $C_{f\!f}$, $C_{bb}$, and $C_{bf}$ to find out where the two species of atoms tend to congregate. 
In case D, the two-body correlation of fermions $C_{f\!f}$ is peaked at the four corners, which means that for a given fermion near a wall, there is a high probability of finding another one near the same or opposite wall. 
In contrast, the boson-boson correlation $C_{bb}$ clearly shows that the bosons occupy the center of the box. Thus, the correlations indicate the system prefers a sandwich structure. 

In case E, the two-body correlation of fermions $C_{f\!f}$ are similar to case A, but the correlations are within a smaller region. This means that the fermions are spread uniformly in the center of the box. Moreover, the boson correlation $C_{bb}$ is now peaked at the four corners, which means that for a given boson near the wall, there is high probability of finding another one near the same or opposite wall. 
The picture of the boson-fermion correlation $C_{bf}$ further corroborates the above interpretation. When the fermions (or bosons) are concentrated in the middle of the box, the bosons (or fermions) will be near the walls. With the analysis, we conclude that the system will show three-chunk phase separation (or sandwich structure) in the many-body limit when imperfections or fluctuations from the atoms or traps are considered.
We remark that for a strong mass imbalance, the heavy particles in a three-chunk structure will occupy the regions near the walls to reduce the kinetic energy due to the distortion of the wavefunctions. 

The two-chunk separation inferred from the two-body correlations can be found in almost all the phase diagrams shown in Fig.~\ref{fig:miscibility}, apart from the one with the smallest mass ratio $m_b/m_f=0.08$. In that special case, the results are limited by the demanding computation to explore the strong-interaction regime, so the appropriate range for two-chunk separation on the phase diagram may not have been covered in our calculation. On the other hand, we did not find three-chunk separation in the correlations of boson-fermion mixtures with comparable masses. While this may be due to the limitation of the parameter space that we can explore, we anticipate the three-chunk separation regime to be small in general, which is consistent with the many-body results that will be shown in the next section.

\section{Many-body Results}\label{sec:manybody}
Here we present the results from many-body mean-field theory of binary boson-fermion mixtures. The first case is with nearly equal masses, exemplified by a mixture of ${}^7$Li and ${}^6$Li, and then cases with larger mass imbalance will be presented, including ${}^7$Li - ${}^{86}$Rb and ${}^6$Li - ${}^{87}$Rb mixtures. Our method is general and applies to other atomic boson-fermion mixtures in 1D box potentials as well.
Unless otherwise specified, we will present the results of $N_b=50=N_f$. We have verified that increasing the particle numbers does not introduce further features.  A $1000$-point grid is used to discretize the space, and we have checked the results are insensitive to a further refinement of the grid. 

\subsection{Comparable masses: ${}^6$Li--${}^7$Li mixture}
After solving the coupled equations of the binary boson-fermion mixtures in a 1D box and comparing the ground-state energies of possible solutions to pick the lowest-energy configuration, we identify the stable ground-state structures of a mixture of ${}^6$Li and ${}^7$Li.

\begin{figure}[t!]
\includegraphics[width=0.9\columnwidth,keepaspectratio]{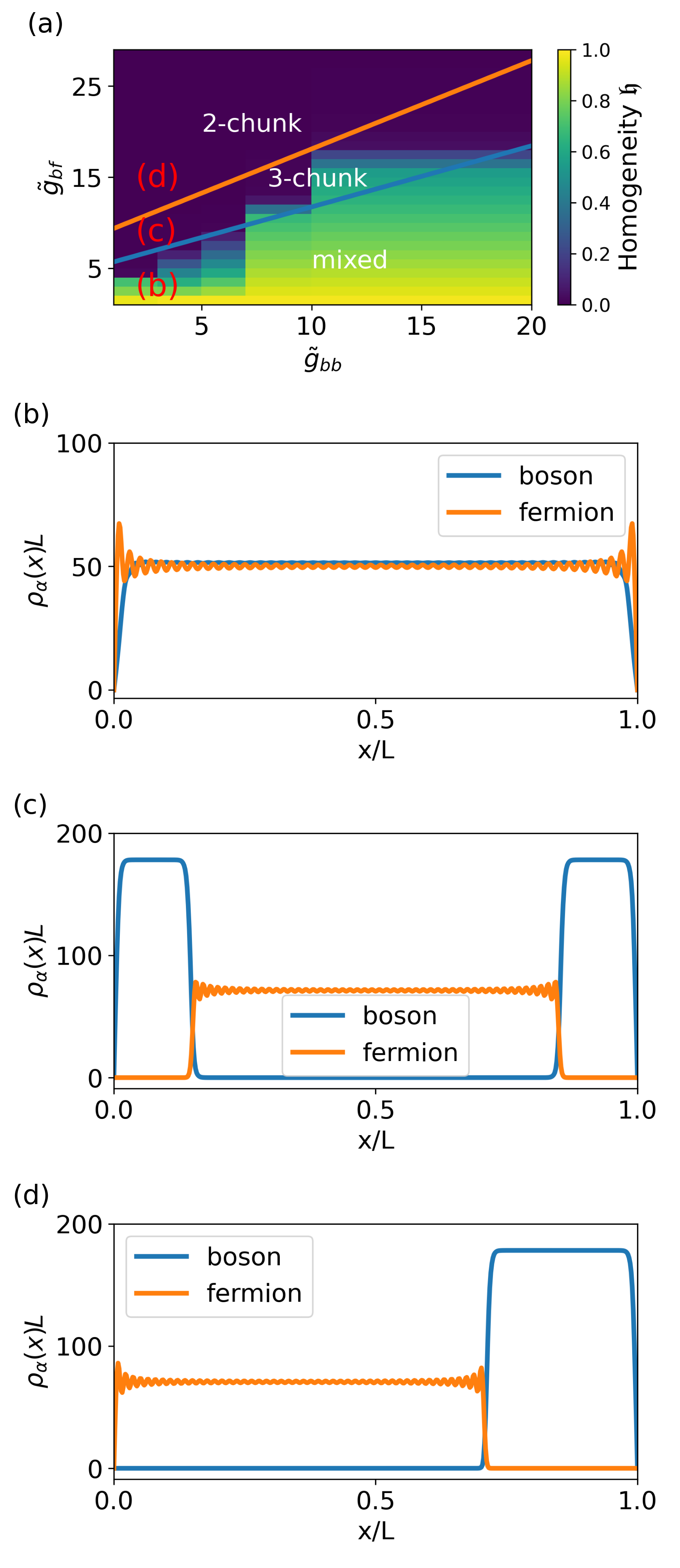}

\caption{\label{box1} Phase diagram (a) and density profiles (b)-(d) of ${}^7$Li--${}^6$Li boson-fermion mixtures with mass ratio $m_b/m_f=7/6$. Here $N_b=N_f=50$ and $\tilde{g}_{bb}=2$ with $\tilde{g}_{bf}=2$ (b), $\tilde{g}_{bf}=8$ (c), and $\tilde{g}_{bf}=14$ (d) with their locations labelled on panel (a). 
}
\end{figure}

\subsubsection{Phase diagram and density profiles}
The $\tilde{g}_{bb}$-$\tilde{g}_{bf}$ phase diagram of the ground-state structures of a mixture of ${}^6$Li and ${}^7$Li is shown in Fig.~\ref{box1}(a). We also show the homogeneity $\mathfrak{h}$, defined in Eq.~\eqref{eq:Homo}.
There are three possible structures: The miscible phase in the weak inter-species interaction region, two-chunk separation in the strongly interacting region, and three-chunk (or sandwich) separation in the intermediate regime. When $\mathfrak{h}\rightarrow 1$, the mixture is in the miscible phase and when $\mathfrak{h}\rightarrow 0$, the bosons and fermions are phase-separated (into either a three-chunk or two-chunk structure). Representatives of the three regimes of Fig.~\ref{box1}(a) are shown in Fig.~\ref{box1}(b), (c), and (d). In the miscible phase, there is a substantial overlap between the two species except the regions near the hard walls, where the wavefunctions are distorted by the boundary condition. In the three-chunk separation, the fermions congregate at the center, enclosed by the bosons on both sides. Finally, in the two-chunk separation, the bosons and fermions occupied opposite sides and break parity symmetry due to imperfections of the initial condition or fluctuations in the calculations. We remark that the total energies of different structures have been compared, and the most stable state is chosen for each set of parameters.

When compared to a previous analysis in an infinitely large system without boundary~\cite{Tom18}, one can see that the three-chunk structure from the mean-field calculation is only possible in the presence of the hard walls. This is because the fermions already have the main contribution to the kinetic energy from the piling-up of the Fermi sea, so they are less sensitive to the distortion at the hard walls. On the other hand, the bosons with finite $\tilde{g}_{bf}$ can have a smoother profile when interfacing with the fermions than with the hard walls. Therefore, the mean-field result of the three-chunk structure in a box potential shows the influence of geometry on quantum systems.
One may observe that the repulsive boson-boson interaction competes with the influence of the repulsive boson-fermion interaction. This is because the condensate of bosons has negligible kinetic energy, so the bosonic self-interaction plays the role of the Fermi pressure and pushes the other species. 

\begin{figure}[t!]

\includegraphics[width=0.9\columnwidth,keepaspectratio]{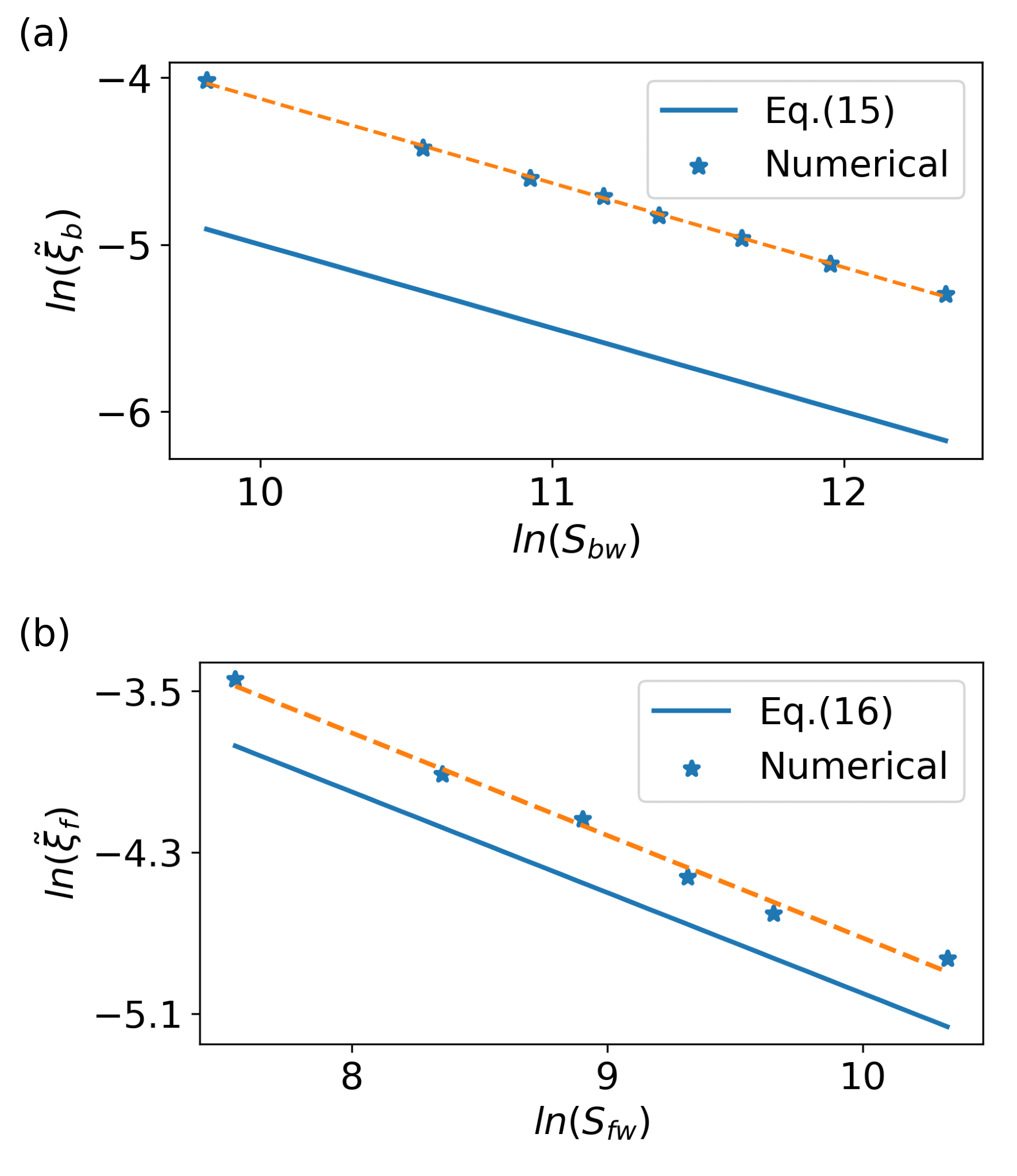}

\caption{\label{xi-walls}Healing lengths of (a) bosons and (b) fermions at the opposite boundaries of the box potential for a ${}^7$Li--${}^6$Li mixture from the simulations and Eqs.~\eqref{eq:xibhw} and \eqref{eq:xifhw}. Here $N_b=N_f=50$  in (a) and (b) with $\tilde g_{bf}=40$ in (a) and $\tilde g_{bb}=1,~\tilde g_{bf}=10$ in (b).}
\end{figure}

On the other hand, a two-chunk structure breaks the parity symmetry in a box potential. If a calculation and its conditions respect the parity symmetry, two-chunk structures will not emerge in the density profile. In our few-body calculations, we analyze the correlations to reveal the underlying two-chunk structure. In our mean-field calculations, however, we use fluctuations in the initial conditions to break the parity symmetry and confirm the two-chunk structure with separating densities becomes the most stable in the strong-interaction regime. For realistic situations in experiments, fluctuations in the preparation, trapping, and manipulations of atoms in the strong-interaction regime may also break the parity symmetry and result in the stable two-chunk structure. 

A closer examination of Fig.~\ref{box1}(b), (c), (d) suggests that increasing the repulsive boson-fermion interaction tends to reduce the width of the overlap between the two species. This is expected because the overlap region incurs high interaction energy. In the following, we will analyze the interface properties of the mixtures.

\begin{figure}[t]
\includegraphics[width=0.9\columnwidth]{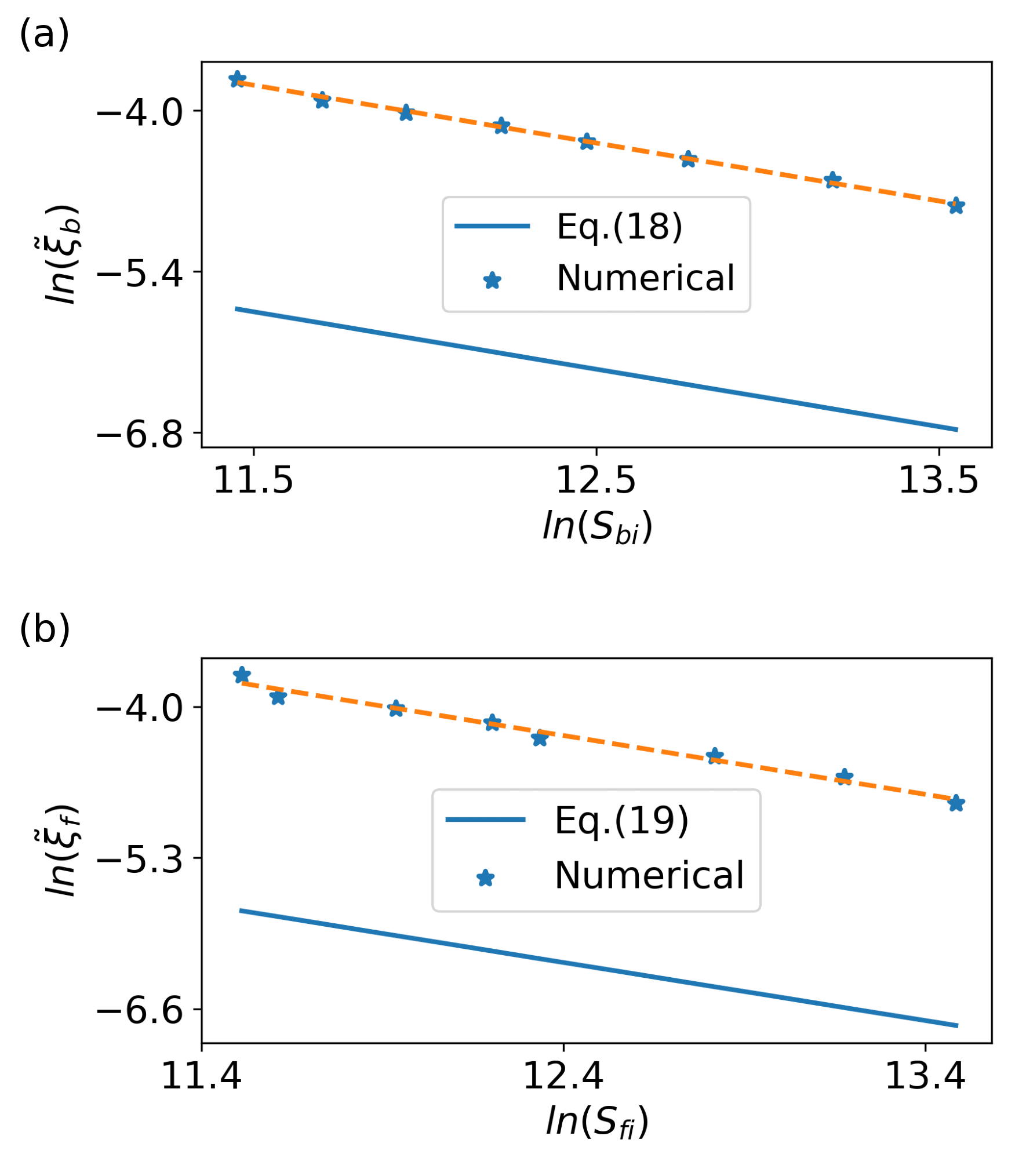}
\caption{\label{xi-interface1}Healing lengths of (a) bosons and (b) fermions at the interface of a ${}^7$Li--${}^6$Li mixture in a two-chunk structure from the simulations and Eqs.~\eqref{xibi} and \eqref{xifi}. 
Here $N_b=N_f=50$ in (a) and (b).}
\end{figure}

\subsubsection{Healing lengths at the walls}
For pure bosons in a box with hard walls, $\psi_b\rightarrow0$ at the walls and $\psi_b$ approaches the constant bulk value away from the boundary. The distance over which the wavefunction rises from zero at the wall to its bulk value is often referred to as the healing (or coherence) length~\cite{Fetter_book,pethick2008bose}. Near the wall, $\psi_b$ is governed by a competition between the kinetic and interaction energies. If we denote the length scale of the variation of the bosons at the wall by $\xi_b$, the
kinetic energy per particle due to the distortion of the wavefunction is given by 
  $KE_b=\frac{\hbar^2}{2m_b\xi_b^2}$.
The healing length of bosons is defined as the length scale at which the kinetic energy per particle matches the interaction energy per particle, $g_{bb}\rho_b$. This leads to an estimation
$\frac{\hbar^2}{2m_b\xi_b^2} \sim g_{bb}\rho_b$. By defining $\tilde \xi_b=\xi_b/L$ as a dimensionless quantity, the scaling of the healing length is
\begin{equation}\label{eq:xibhw}
    \tilde \xi_b \sim \sqrt{\frac{m_f}{m_b k_f^0L}}\frac{1}{\sqrt{\tilde  g_{bb} \rho_b L}}\equiv \frac{1}{\sqrt{S_{bw}}}.
\end{equation}
Here we define a dimensionless parameter $S_{bw}\equiv \tilde{g}_{bb}\rho_b k_f^{0} L^2 m_b/m_f$ to simplify the scaling analysis. The presence of the fermionic parameters is only to fix the units.

Meanwhile, the length scale of the variation of fermions at the wall, denoted by $\xi_f$, may be determined by matching the kinetic energy per particle with the Fermi energy. The reason is because Pauli exclusion principle may be viewed as an effective (statistical) interaction between fermions, leading to an energy scale determined by the Fermi energy $E_f$. The balance $KE_f=\hbar^2/(2m_f \xi_f^2) \sim E_f$ then leads to $\xi_f \sim \frac{1}{k_f}$. Here $k_f$ is the bulk Fermi wavevector, determined by the bulk fermion density $\rho_f$ via $k_f=\pi\rho_f/2$. In terms of the dimensionless healing length $\tilde\xi_f=\xi_f/L$, we have
\begin{equation}\label{eq:xifhw}
    \tilde \xi_f \sim \frac{1}{k_f L}\equiv \frac{1}{\sqrt{S_{fw}}}.
\end{equation} 
Here we define another dimensionless parameter $S_{fw}\equiv (k_f L)^2$ to simplify the scaling analysis.

Now we consider two-chunk separation in the strong-interaction regime, where the bosons occupy one side of the box while the fermions occupy the other side. In such configurations, there is practically only one species near each hard wall.
In our analyses of the healing lengths, we take the width as the distance between $95\%$ and $5\%$ of the value of $\sqrt{\rho_\alpha}$ at the plateau in the bulk. Taking different criteria or using functional fits to the density profiles leads to basically the same scaling behavior, which verifies the robustness of the energy-competition argument.
Fig.~\ref{xi-walls} shows the scaling of the healing lengths of the bosons and fermions near the hard walls, respectively. The scaling behavior confirms the arguments based on the competition of the kinetic and interaction energies for each species. We note that the energy-competition arguments do not fix the pre-factors of the healing lengths, causing a parallel shift between the data and analytic formulas on a log-log plot.

\subsubsection{Healing lengths at the interface}
In the phase-separation structures, both species are present at the interface between the two species. The interaction energy (IE) per particle of the bosons and fermions at the boson-fermion interface may be respectively estimated as
\begin{eqnarray}
    IE_b=g_{bb}\rho_b+g_{bf}\rho_f,~ IE_f=E_f + g_{bf}\rho_b.
\end{eqnarray}
For the fermions, Pauli exclusion principle may be considered as an effective (statistical) interaction, which introduces the Fermi energy $E_f$ to $IE_f$.
If $\xi_\alpha$ denotes the healing lengths for species $\alpha=b, f$, then the kinetic energy per particle due to the distortion of the wavefunction is again given by $KE_\alpha=\frac{\hbar^2}{2m_\alpha\xi_\alpha^2}$.
As discussed earlier, the healing lengths may be estimated using the conditions $KE_\alpha\approx IE_\alpha$. Explicitly, for the bosons,
$\frac{\hbar^2}{2m_b\xi_b^2}\sim  g_{bb}\rho_b+ g_{bf}\rho_f$,
which leads to
\begin{equation}\label{xibi}
\tilde{\xi}_b\sim \sqrt{\frac{m_f}{m_b k_f^0L}} \frac{1}{\sqrt{ \tilde g_{bb}\rho_b L+ \tilde g_{bf}\rho_f L}}\equiv \frac{1}{\sqrt{S_{bi}}}.
\end{equation}
For the fermions,
    $\frac{\hbar^2}{2m_f\xi_f^2}\sim \frac{\hbar^2 k_f^2}{2m_f}+g_{bf}\rho_b$,
which leads to
\begin{equation}\label{xifi}
\tilde \xi_f\sim \frac{1}{\sqrt{(k_f L)^2+\tilde g_{bf}(\rho_b L)(k_f^0 L)}}\equiv \frac{1}{\sqrt{S_{fi}}}.
\end{equation}
Similar to the analyses of the healing lengths at the hard walls, here we define two dimensionless parameters $S_{bi}\equiv[\tilde{g}_{bb}\rho_b L +\tilde{g}_{bf}\rho_f L]m_b k_f^0 L/m_f$ and
$S_{fi}\equiv [(k_f L)^2+\tilde{g}_{bf}\rho_b k_f^{0} L^2]$ to simplify the scaling analyses of the healing lengths at the boson-fermion interface. We note that when comparing the analyses of the healing lengths at the hard walls versus those at the boson-fermion interface, the expressions of $S_{\alpha i}$ for $\alpha=b, f$ are consistent with those of $S_{\alpha w}$ because only one species is present near each hard wall in two-chunk phase separation but both species are present at the boson-fermion interface.

We remark that in the expressions of the healing lengths, $\rho_\alpha$ denotes the bulk density of the corresponding species away from the interface or hard wall, and $k_f^0=\pi N_f/2L$ is the noninteracting Fermi wavevector while $k_f=\pi\rho_f$ is the bulk Fermi wavevector of the fermions in the mixture. The interface widths for both species from the simulation results can be obtained from the density profiles by following the same analyses as we did for the healing lengths at the hard walls. Moreover, we have verified that taking different criteria or using functional fits to the density profiles basically leads to the same scaling behavior.

Fig.~\ref{xi-interface1} shows that the healing lengths of the bosons and fermions at the interface scale according to Eqs. (\ref{xibi}) and (\ref{xifi}), respectively, in the two-chunk regime shown in Fig.~\ref{box1}. Since testing the scaling behavior requires a broad range of parameters, the two-chunk regime is more appropriate because the three-chunk regime is narrow along the $\tilde{g}_{bf}$ direction. The agreement of the scaling behavior between the simulations and analytical formulas of the healing lengths verifies that the energy-competition argument works well with binary boson-fermion mixtures in a 1D box. We remark that more complicated analyses with constructions of piecewise energy functionals~\cite{interface,surftension,normal-superfluid} may lead to refinements of the structures and interface widths, which will in turn determine the pre-factors of the healing lengths that cannot be explained by the scaling analysis. Nevertheless, the simple scaling from energy-competitions provides us the main physical meaning for explaining future experiments on atomic boson-fermion mixtures.

\subsection{${}^{87}$Rb-${}^{6}$Li: heavy bosons and light fermions}

\begin{figure}[t]

\includegraphics[width=0.9\columnwidth]{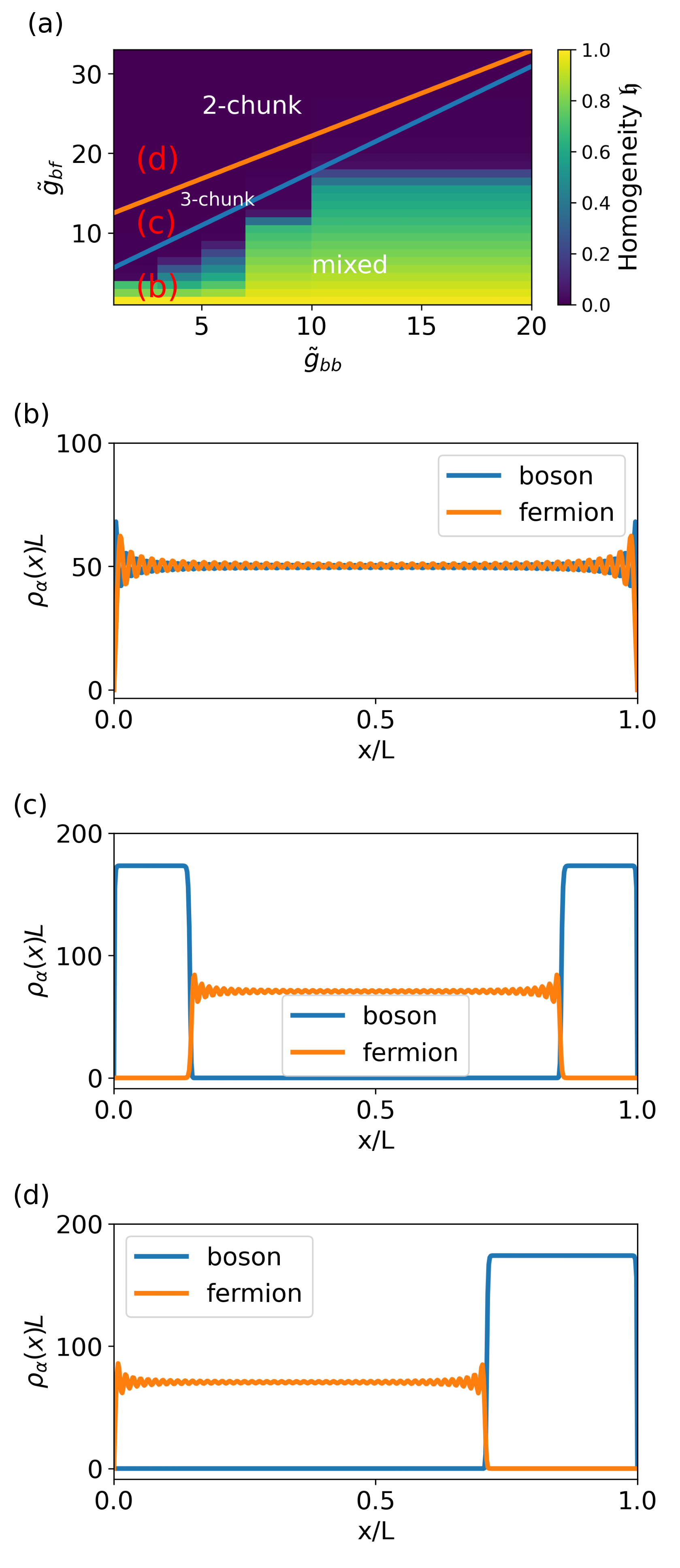}
    
\caption{\label{box2} Phase diagram (a) and density profiles (b)-(d) of ${}^{87}$Rb--${}^6$Li mixtures. Here $N_b=N_f=50$ and $\tilde{g}_{bb}=2$ with $\tilde{g}_{bf}=2$ (b), $\tilde{g}_{bf}=10$ (c), and $\tilde{g}_{bf}=18$ (d) with their locations labelled on panel (a). 
}
\end{figure}

\subsubsection{Phase diagrams and structures}
For boson-fermion mixtures with prominent mass imbalance, we first analyze mixtures of ${}^{87}$Rb and ${}^{6}$Li. This system also exhibits the miscible phase, three-chunk separation, and two-chunk separation of the two species as the inter-species interaction increases. Fig.~\ref{box2}(a) shows the phase diagram of ${}^{87}$Rb and ${}^{6}$Li mixtures. Fig. \ref{box2}(b), (c), and (d) show the representative density profiles of the miscible phase in the weak inter-species interaction regime, three-chunk (sandwich) separation in the intermediate interaction regime, and the two-chunk separation in the strong interaction regime, respectively. Heavier mass lowers the kinetic energy  due to distortion of the density profile in a phase-separation structure because the mass appears in the denominator of the kinetic energy. To minimize the kinetic energy due to the distortion of the wavefunction in the three-chunk structure, the density of the lighter species tends to stay away from the hard walls while the heavier species tends to occupy the region there until the two-chunk structure becomes energetically more favorable than the three-chunk structure. 

\begin{figure}[t]

\includegraphics[width=0.9\columnwidth,keepaspectratio]{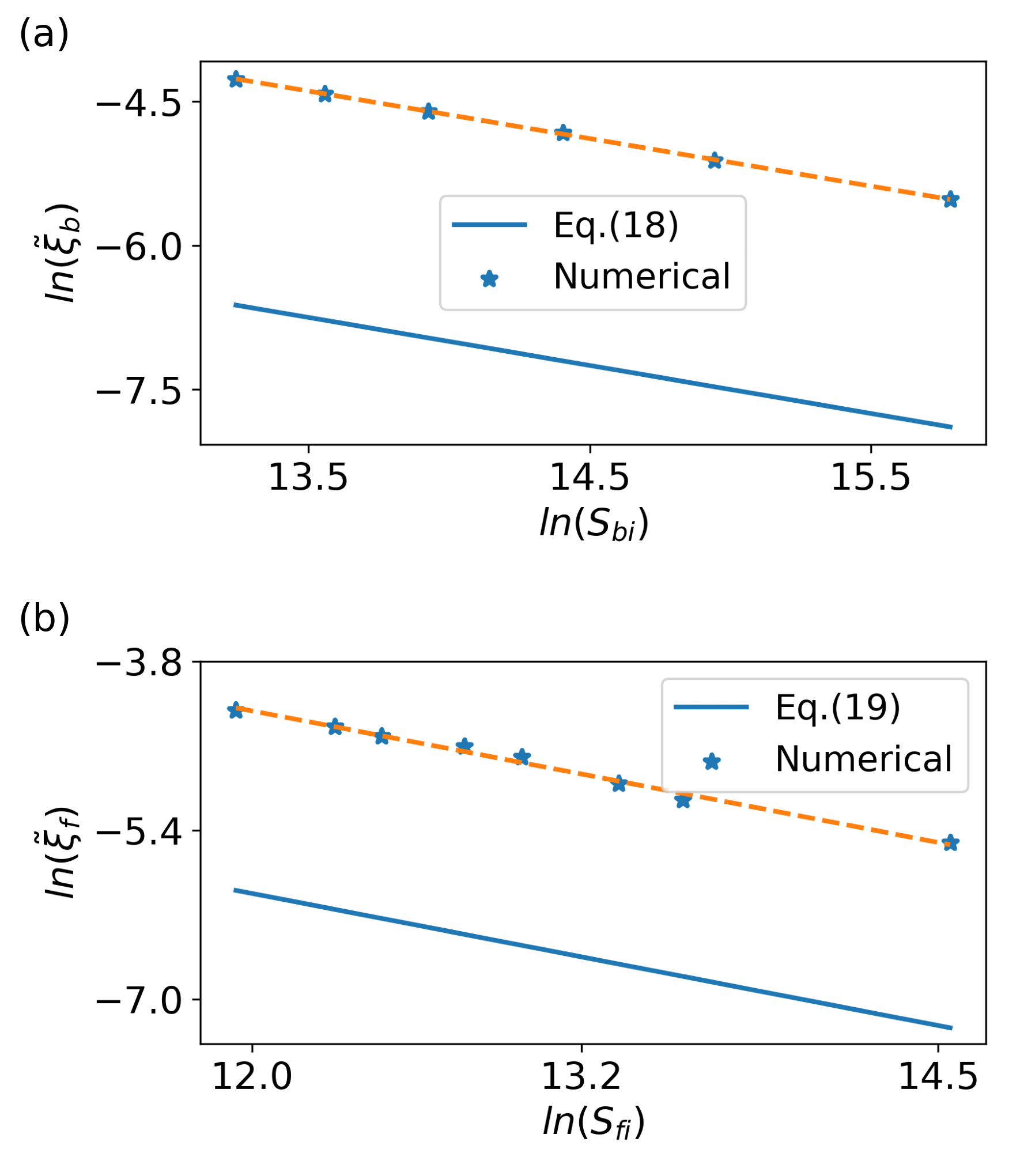}
\caption{\label{xi-interface2} Scaling behavior of the healing lengths of (a) bosons and (b) fermions at the interface of a ${}^{87}$Rb - ${}^{6}$Li mixture in the two-chunk separation from the simulations and Eqs.~\eqref{xibi} and \eqref{xifi}. Here $N_b=N_f=50$ in (a) and (b). }
\end{figure}

\subsubsection{Interface properties of ${}^{87}$Rb - ${}^{6}$Li mixture }
The healing lengths of the bosons and fermions with larger mass imbalance at the hard walls in a 1D box are found to follow the same scaling as their counterparts in the ${}^7$Li--${}^6$Li mixture. Hence, we do not repeat the analysis of $\xi_\alpha$ at the hard walls. For the interface between the two species, we found that since the three-chunk structure exists in a narrow parameter range, it is more challenging to analyze the scaling behavior. Therefore, the interface properties of the mixtures are discussed only for the two-chunk structure that extends well into the strongly interacting regime.

For $^{87}$Rb - $^{6}$Li mixtures,  we found that the healing lengths of the bosons and fermions at their interface scale according to Eqs. (\ref{xibi}) and (\ref{xifi}), respectively. The scalings of the healing lengths with interactions are shown in Fig. \ref{xi-interface2}, which confirm that the widths of the bosons and fermions at the interface are determined by competitions between the kinetic energy due to the distortion of the density and the interaction energy, which includes the inter- and intra- species interactions and the effective (statistical) interaction of fermions. Furthermore, the scaling analyses correctly capture the power-law dependence of the healing lengths of the bosons and fermions at the phase-separation interface and verify the energy-competition argument.

\begin{figure}[t]

\includegraphics[width=0.9\columnwidth]{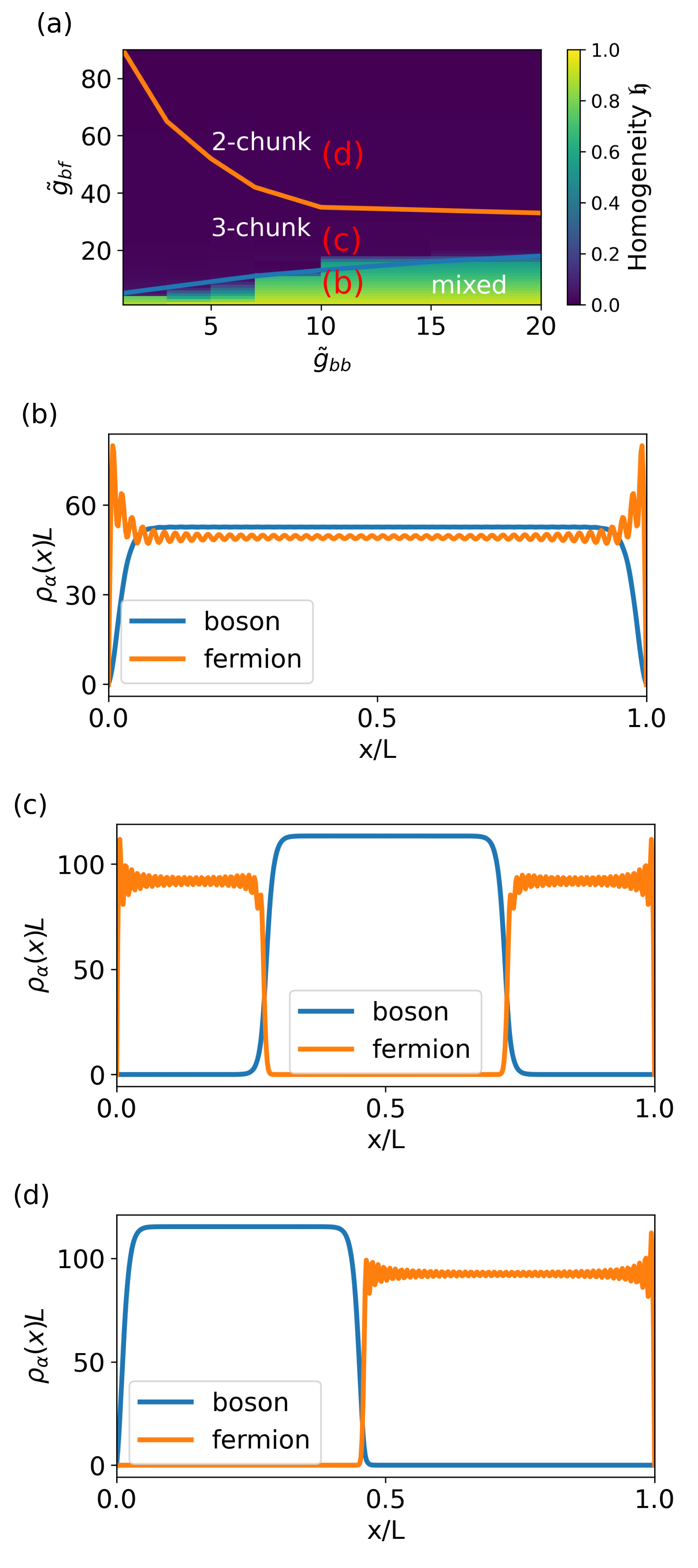}

\caption{\label{box3} Phase diagram (a) and density profiles (b)-(d) of ${}^{7}$Li--${}^{86}$Rb boson-fermion mixtures. Here $N_b=N_f=50$ and $\tilde{g}_{bb}=10$ with $\tilde{g}_{bf}=5$ (b),  $\tilde{g}_{bf}=20$ (c), and $\tilde{g}_{bf}=50$ (d) with their locations labelled on panel (a).} 
\end{figure}

\subsection{${}^{7}$Li-${}^{86}$Rb: light bosons and heavy fermions}

\subsubsection{Phase diagram and structures}
Here we consider a ${}^7$Li-${}^{86}$Rb mixture as an example of light bosons and heavy fermions. The phase diagram is shown in Fig.~\ref{box3}(a) with a relatively large three-chunk regime when the boson-boson  interaction is weak. Interestingly,  we could not reach the regime in the few-body calculations to observe two-chunk separation in ${}^{7}$Li-${}^{86}$Rb mixture due to the demanding computation, which implies that the parameter space for the three-chunk structure is relatively large. This is indeed the case from the many-body result. 

For the three-chunk structure, the boson is now in the center of the box because the kinetic energy is relatively small for the heavy fermions, which tend to stay near the hard walls and push the bosons away from the hard walls. Meanwhile, the bosons rely on the boson-boson interaction to build up pressure to push against the fermions. Hence, the three-chunk structure remains energetically favorable than the two-chunk structure when $\tilde{g}_{bb}$ is weak and the bosons cannot repel the fermions at both hard walls.  When $\tilde{g}_{bf}>>\tilde{g}_{bb}$, however, the bosons are tightly compressed by the fermions and eventually pushed to one side of the box to form a two-chunk structure with lower total energy.

\begin{figure}[t]

\includegraphics[width=0.9\columnwidth,keepaspectratio]{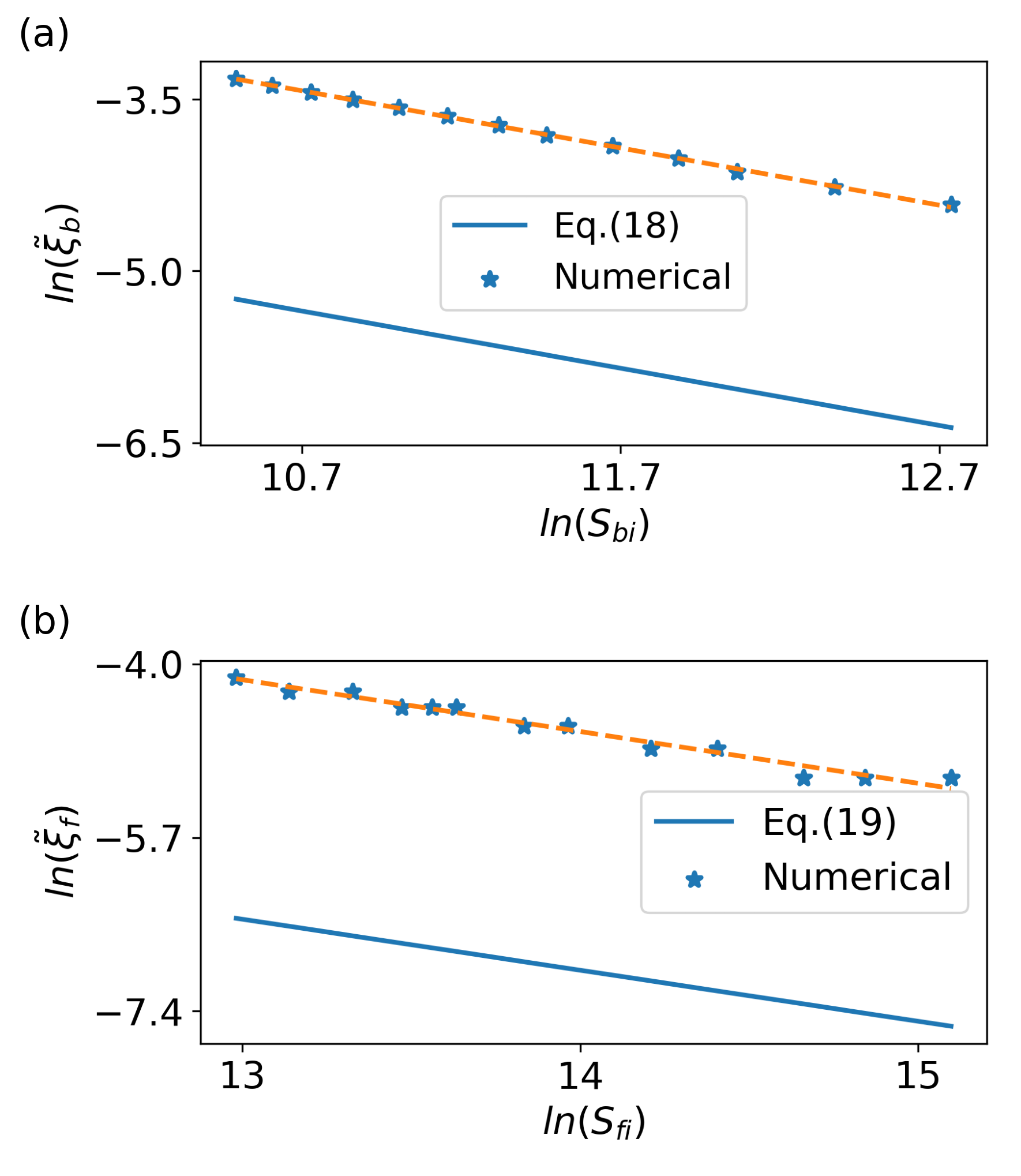}

\caption{\label{xi-interface3}Healing length of (a) bosons and (b) fermions at the interface of a $^{7}$Li - $^{86}$Rb mixture in the two-chunk separation from the simulations and Eqs.~\eqref{xibi} and \eqref{xifi}. Here $N_b=N_f=50$ in (a) and (b). }
\end{figure}

\subsubsection{Interface properties of ${}^{7}$Li - ${}^{86}$Rb mixture}
Since the density profiles of the ${}^{7}$Li - ${}^{86}$Rb mixture in the two-chunk regime are similar to those of the ${}^{87}$Rb - ${}^{6}$Li mixture, the analyses of the widths that reflect the healing lengths at the phase-separation interface are also similar. We again focus on the two-chunk regime due to its broad coverage of the strong-interaction region on the phase diagram. The scaling of the healing lengths with the interactions are shown in Fig.  \ref{xi-interface3}. The same scaling analysis confirms the functional dependence of the healing lengths of the bosons and fermions described by Eqs.~\eqref{xibi} and \eqref{xifi} at the phase-separation interface, showing the generality of the energy-competition argument. We mention that the scaling analyses only reveal the functional forms of the healing lengths, and the pre-factors need to be determined from simulations or direct evaluations of the energy functionals for the inhomogeneous systems. Moreover, we have verified the scaling of the healing lengths with more particle numbers or grid points, and the functional forms remain the same because of the energy-competition mechanism.

\begin{figure}[t]

\includegraphics[width=0.99\columnwidth,keepaspectratio]{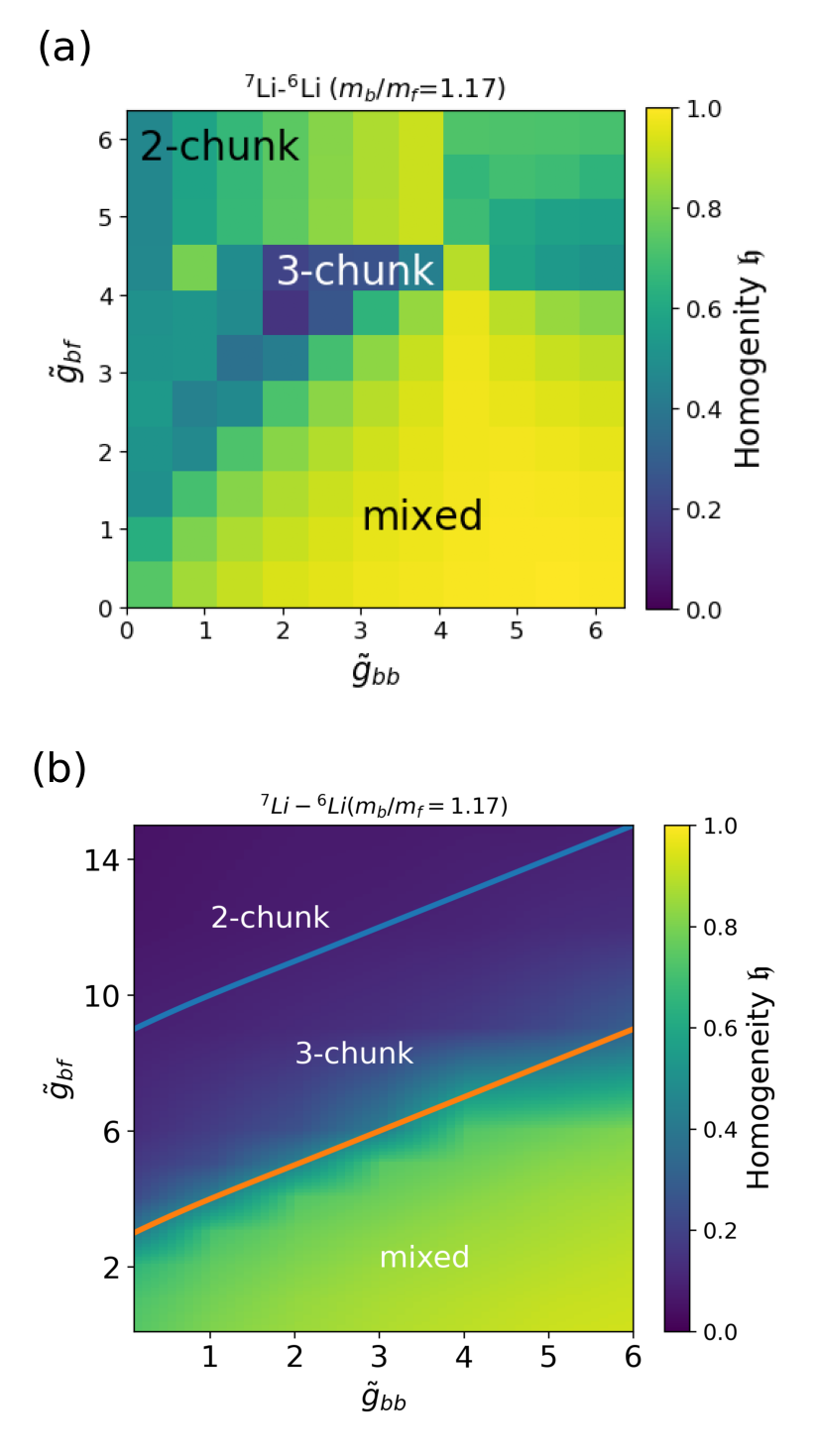}

\caption{\label{fig:comparison} Phase-diagrams of ${}^7$Li - $^6$Li boson-fermion mixtures from (a) few-body calculation and (b) mean-field treatment for $N_b=4=N_f$.} 
\end{figure}

\section{Implications}\label{sec:implications}
We verify the qualitative agreement between the few-body exact calculation and mean-field treatment by comparing the phase diagrams of $^6$Li - $^7$Li mixtures in Fig. \ref{fig:comparison}. To facilitate a fair comparison, we use $N_b=4=N_f$ for the mean-field treatment as well. As one can see in both diagrams, the miscible phase starts to separate as $\tilde{g}_{bf}$ increases, with a three-chunk regime in between the miscible phase and the two-chunk regime. The two approaches indeed agree qualitatively despite some subtle differences in the details.

To analyze the interface structures of atomic mixtures, high-resolution imaging of cold-atom systems at the level of single-atom sensitivity is desired. Microscopy techniques inspired by the scanning electron microscopy have been successfully used for detection of single atoms inside a quantum gas in an optical lattice with a resolution better than $150~ nm$~\cite{SEM-high-res} but with low accuracy of single-atom detection. The invention of quantum gas microscopes, for example, allows sub-micron resolution of the order of $0.5~\mu m$ and near-unity detection efficiency by fluorescence imaging in a pinning lattice for ultracold atoms in optical potentials~\cite{Bakr_2009}. A review of recent developments of quantum gas microscopes is given in Ref.~\cite{GrossChristian2021Qgmf}. Super-resolution imaging based on nonlinear response of atoms in an optical lattice to an optical pumping is demonstrated with the capability of imaging the structure of an atomic cloud with a resolution of $30$nm and a localization precision of the pinning lattice below $500~pm$~\cite{PhysRevX.9.021001}. 
Recently, trapped few-body systems with experimentally relevant setups have been analyzed by using the quantum point spread function~\cite{qpsf-quantum-microscope}, suggesting a resolution of $(1/33) a_l$ smaller than the lattice spacing ($a_l$) of the pinning lattice. 

Those progresses will guide experimental efforts towards high-resolution imaging in non-lattice systems as well. For a typical 1D box potential of length $L=160 \mu m$~\cite{Tajik:19}, the resolution required to analyze the interface width of $\sim L/100\sim1.6\mu m$ for studying the healing lengths is reasonably within the experimental resolution limit. With the rapid developments of cold-atom microscopes capable of higher resolution beyond lattice systems, the widths of different atomic species at the interface will reveal interesting thermodynamic properties of interacting multi-species quantum systems via their structures.

Furthermore, we remark that atomic mixtures like the ${}^{7}$Li - ${}^{86}$Rb mixtures are promising to track down the elusive signatures of the so-called Fulde-Ferrell-Larkin-Ovchinnikov (FFLO) state, which has been predicted in two-component fermionic mixtures with attraction and population imbalance~\cite{FF64, LO65}. The conventional Cooper mechanism cannot be fulfilled in the presence of high population imbalance, and the Cooper pairs may posses non-vanishing net momentum to maximize the pairing between the two components of the fermions in the FFLO state. There have been experimental evidence suggesting the FFLO state in quasi-1D two-component ${}^{6}$Li gases with population imbalance~\cite{Liao10}. Few-body calculations suggest the FFLO state may be enhanced by suitable confinement~\cite{Dobrzyniecki2021Unconventional}. To directly observe the FFLO state, one has to measure the two-body correlations \cite{Pecak2020FFLO, pecak2022unconventional}, which may be inferred in state-of-the-art experiments.

However, standard experiments usually involve single-particle measurements, for example, measurements of the density.
It has been shown \cite{singh2020enhanced} that the FFLO correlations might be strongly enhanced by interactions with bosons. Since three-chunk structures naturally occur due to interactions in boson-fermion mixtures confined in 1D box potentials, such settings provide a feasible route for enhancing and probing the FFLO correlations. 
Nevertheless, the system for observing the FFLO state will involve three components: one species of bosons repelling both components of fermions, and two-component fermions with population imbalance and attractive interactions.
Since the $^{7}$Li - $^{86}$Rb  mixture has a broad range of the three-chunk regime as shown in our analysis, strong correlations and Pauli exclusion principle of the fermions may immune the system from a collapse into a two-chunk structure when two internal states of the fermions with attractive interactions are introduced, which may provide a feasible setup for future studies of the FFLO state.

\section{Conclusion}\label{sec:conclusion}
We have presented both few-body calculations and many-body mean-field approximations of binary atomic boson-fermion mixtures in 1D box potentials, showing different structures as the parameters vary. The stable structures are determined by competitions between the interaction and kinetic energies, which are further complicated by the presence of the hard walls, mass imbalance, and boson-fermion interface if the two species separate.
While the few-body results reveal the correlations among the bosons and fermions, the many-body results allow a systematic extraction of the healing lengths of the bosons and fermions. Moreover, the scaling behavior of the healing lengths at the boson-fermion interface in the phase-separation regime confirms the energy-competition mechanism behind the structures of binary atomic boson-fermion mixtures. With advancement in trapping and manipulating atoms in box potentials, our results may be verified in coming experiments and inspire future research.

\begin{acknowledgments}
We thank Prof. Chien-Te Wu for stimulating discussions. C. C. C. was supported by the National Science Foundation under Grant No. PHY-2011360.
D. P. was supported by the (Polish) National Science Center Grants No. 2021/40/C/ST2/00072.
\end{acknowledgments}

\bibliographystyle{apsrev}

\end{document}